\documentstyle[preprint,aps]{revtex}
\begin{document}
\draft
\title{The Central Temperature of the Sun can be Measured via the
$^7$Be Solar Neutrino Line}

\author{John N. Bahcall}

\address{Institute for Advanced Study, Princeton, NJ 08540}

\date{\today}

\maketitle

\begin{abstract}
A  precise test of the theory of stellar
evolution can be performed by measuring the difference
in average energy between the neutrino line
produced by ${\rm ^7Be}$ electron capture in the
solar interior and the corresponding neutrino line
produced in a terrestrial laboratory.
The high temperatures in the center of the sun broaden
the line asymmetrically, FWHM = 1.6~keV, and cause an
average energy shift of 1.3~keV. The
width of the $^7$Be neutrino line should be taken into
account in calculations of vacuum neutrino oscillations.

\end{abstract}


The purpose of this letter is to describe the observational implications
of a precise,
new prediction of the theory of stellar evolution and to stimulate
thinking about possible experimental tests. The
predicted quantity is the energy profile of the neutrino line
produced by ${\rm ^7Be}$ electron capture
in the solar interior.
Standard solar models imply  that the average energy of the
$^7$Be neutrino line profile is shifted
from its laboratory value by 1.3~keV and is broadened to a full-width at
half-maximum of 1.6~keV.
The predicted energy shift may be measurable with
future solar neutrino experiments.
The details of the calculations that lead to these results will be given
elsewhere\cite{Bahcall93}; only the physical ideas and the principal
observational consequences
will be summarized here (for background material,
see\cite{Bahcall89}).

The reaction in question is
\begin{equation}
e^- + ^7{\rm Be}\  \to\ ^7{\rm Li} + \nu_e\ .
\label{capturereaction}
\end{equation}
Reaction (\ref{capturereaction}) occurs
in about 15\% of the terminations of the
proton-proton chain fusion reactions in standard solar
models.
In the interior of the sun, most
of the electrons are captured from
continuum states.
The reaction produces a line because the recoiling nucleus takes up a
significant amount of momentum but only a negligible amount of energy.
I consider here only the experimentally more-accessible transition to the
ground state of $^7$Li.

The average energy of the solar neutrino line is larger
than for the corresponding laboratory decay because electrons and ${\rm ^7Be}$
ions move with appreciable kinetic energies at the high
temperatures ($\approx 1$~keV) characteristic of the solar interior.
The average energy difference between neutrinos emitted in solar and in
laboratory decays reflects the temperature profile in the center of the
sun and is determined primarily by the center-of-momentum thermal energy of
the continuum electrons and of their capturing ${\rm ^7Be}$ nuclei,
by the Doppler shifts of
the ${\rm ^7Be}$ ions, by the fraction
of electron captures that occur from
bound orbits rather than from continuum orbits, and by the difference in
atomic binding energies between solar and laboratory conditions.

The average shift, $\Delta$, in the neutrino energy, $q$,
between $^7$Be neutrinos emitted in the sun and $^7$Be neutrinos emitted
in the laboratory is\cite{Bahcall93}

\begin{equation}
\Delta\ \equiv \Bigl\langle q - q_{\rm lab}\Bigr\rangle = \left(1.29
 \pm 0.01 \right)\ {\rm keV}
\label{Deltadefinition}
\end{equation}
for three precise, recently-calculated\cite{Bahcall88} solar models
(constructed with and without including helium diffusion).

The energy shift, $\Delta$, is approximately equal to
the average temperature of the
solar interior weighted by the fraction of $^7$Be neutrinos that are
produced at each temperature\cite{Bahcall93}, \hbox{i.e.}
$\int_\odot dTT d\phi({\rm ^7Be}, T)/\int_\odot dT d\phi({\rm
^7Be}, T)$, where $d\phi({\rm ^7Be},T)$ is the flux of $^7$Be
neutrinos produced at the temperature $T$.
The $^7$Be neutrinos are
produced in the inner few percent of the solar mass,
75\% in the region $\left(0.04 \pm 0.03 \right)M_\odot$.
Therefore,  a measurement of the energy shift is a
measurement of the first moment of the
central temperature distribution of the sun.

Figure~I shows the calculated line profile derived by
including the known physical effects.
The neutrino energy at the profile peak is
$q_{\rm peak} = 862.27~{\rm keV}$; the laboratory energy is
$q_{\rm peak} = 861.84~{\rm keV}$.
The line profile is asymmetric.
Doppler shifts caused by thermal velocities of the $^7$Be nuclei
symmetrically broaden the line and
determine the Gaussian shape below the peak.
The low-energy side of the line profile,
$q_{\rm obs} < q_{\rm peak}$,
is produced by $^7$Be nuclei that are moving away from the observer.
The profile at higher energies is determined by the
center-of-momentum kinetic energies:
\hbox{${\rm Spectrum}_{\rm solar}\left(q_{\rm obs}\right)\ \propto\
 \exp\left[-\left(q_{\rm obs} - q_{\rm peak}\right)/kT_{\rm eff}\right].$}
Since kinetic energies are always positive, the exponential tail is
present only for \hbox{$\ q_{\rm obs} > q_{\rm peak}$}.

On the
low-energy side, the standard solar model with helium
diffusion\cite{Bahcall88} predicts a half-width
at half-maximum of 0.56~keV, and on the high-energy side,
the half-width at half-maximum
is 1.07~keV.  The effective temperature of the
high-energy exponential tail is $15 \times 10^6$~K.
The effects of the electrostatic energy of the screening charge
around the $^7$Be and $^7$Li nuclei,
of gravitational redshifts on the neutrinos,
and of collisional broadening of the line
are much smaller than the direct effects of the thermal kinetic energies and
Doppler shifts.

The precision with which the central thermal structure of the sun is
determined by standard solar models and the fundamental and unique
character of the
prediction justifies a special experimental
effort to measure $\Delta$.
For the standard solar
models computed by the author and his colleagues
over the past decade,
the central temperature has varied over a total range of $\pm 0.5$\%,
$T_c = (15.58 \pm 0.08) \times 10^6~{\rm K}$.
For a heterogeneous set of nine recently-calculated
solar models\cite{Berthomieu93}, computed by different groups for
different purposes using different input data and
generally not required to
have the highest-attainable precision, the central temperature varied by
$\pm 1$\%, $T_c = (15.55 \pm 0.15) \times 10^6~{\rm K}$.

Detailed calculations show\cite{Bahcall93} that the characteristic modulation
of the shape of the ${\rm ^7Be}$ neutrino line that would be
caused either by vacuum neutrino
oscillations or by
matter-enhanced (MSW) neutrino oscillations is small.
Other frequently-discussed weak interaction solutions to the solar
neutrino problem, such as
rotation of the neutrino magnetic moment, matter-enhanced
magnetic moment transitions, and neutrino decay,
will also not change significantly the line
profile.  The basic reason for the smallness of all these effects is that
the ratio of the width of the line to the typical neutrino
energy is only $\approx
0.001$, although the influence of fine-tuning must also be calculated.

The energy profile of the $^7$Be neutrino line should be included
in precise calculations
of what is expected from
vacuum neutrino oscillations.  It has become standard\cite{Krastev93}
to take account of the variation of the
distance between the point of creation of the neutrinos
and the point of detection.  The variation in the point of
creation corresponds to a
phase-change, $\delta \phi$,
of order $10^{-4}$, in the phase-angle, $\phi$, that determines the
probability of observing an electron-type  neutrino on earth
(probability $\propto \sin^2\phi$).
(The ratio of the
solar radius to the earth-sun distance is about $0.005$ and
$^7$Be neutrinos are produced in a region of about $\pm 0.025
R_{\odot}$.)
Therefore, the change in phase, $\approx 10^{-3}$,
due to the energy-width of the neutrino
line is an order of magnitude larger than the phase-change caused
by averaging over the region of production (and is comparable to the
phase change due to seasonal variations).

A number of experiments have been proposed that
would measure the $^7$Be neutrino flux with
detectors that are based upon neutrino-electron
scattering\cite{Raghavan90,Drukier84}.
Radiochemical\cite{Rowley78,Haxton88} and electronic
detectors\cite{Raghavan76} of the total $^7$Be neutrino flux have also
been proposed.
The BOREXINO experiment\cite{Raghavan90}
is the most advanced of these proposals and can, if recent estimates of
backgrounds are correct, measure the total flux of $^7$Be neutrinos.
The $\nu-{\rm e}$ scattering experiments will probably not be able to
measure $\Delta$ since for most scattering events
the neutrino and the electron share the final state energy
(which is much larger than $\Delta$, see Figure 8.5 of
\cite{Bahcall89}).  Some experiment that measures the total flux at
earth in the $^7$Be neutrino line should be performed before the
experiments proposed here are attempted, since the total flux may be
reduced (with respect to the prediction of the standard solar model) by
neutrino oscillations or by other new physics.

The most direct way to study the $^7$Be energy profile may be to
detect neutrino absorption by nuclei, which leaves an electron and a
recoiling nucleus in the final state.
Nearly all of the
initial neutrino energy is transferred to the final-state electron
(the nuclear recoil energy being small).

Lithium detectors, which have been discussed\cite{Rowley78} as
attractive detectors of solar neutrinos,
deserve further study since the absorption cross
section for this special case depends sensitively upon $\Delta$.
The calculated absorption cross section for the reaction
\hbox{$\nu_e + {\rm ^7Li} \to {\rm ^7Be} + e^-$},
where $\nu_e$ is produced by $^7$Be electron capture in the sun, depends
upon the assumed energy profile of the solar neutrinos.
Neutrinos cannot be absorbed in this reaction if their energies lie
below the energy threshold of 861.96~keV.
The location of the threshold
within the line profile determines the fraction of
emitted neutrinos that can be absorbed.
The absorption cross section for solar-produced $^7$Be neutrinos
incident on a laboratory detector of $^7$Li is\cite{Bahcall93}
\begin{equation}
\langle{\rm Spectrum}_{\nu_e}(q_{\rm obs})\ \sigma_{\rm abs}(q_{\rm
obs})\rangle\simeq 19 \times 10^{-46}~{\rm cm}^2 ,
\label{crosssection}
\end{equation}
assuming neutrinos do not change flavor after their creation.
As usual, Eq.\ (\ref{crosssection}) includes a correction
for the fact that only 89.7\% of the $^7$Be neutrinos are
produced in ground-state to ground-state transitions.
The cross section given in Eq.\ (\ref{crosssection}) is almost a factor
of two larger than obtained previously \cite{Domogatsky69},
which should make the contemplated experiments somewhat
easier than previously considered.
The earlier treatments neglected the difference in electron binding
energies of solar and laboratory $^7$Be atoms as well as Doppler shifts of
the $^7$Be nuclei, and did not average over the temperature profile of
the sun.
If the flux of $^7$Be electron-type neutrinos were measured in an independent
experiment, the total absorption rate in a lithium detector could be
used to determine $\Delta$.

For absorption detectors, am energy
resolution (${\Delta E}/E$) of order $0.1\%$ to $1.0\%$ is desirable
to measure the 1.3~keV (0.15\%) energy shift.
Detectors have been developed\cite{Caldwell88} for a variety of
applications, including dark matter searches, the observation of double
beta decay, and x-ray astronomy, that have energy resolutions
of better than 1\%, but their surface areas are not yet large enough
for a practical solar neutrino detector.
Consider,
for specificity, a conceivable cryogenic experiment\cite{Alessandrello92}
that might be performed on $^{81}$Br with an energy resolution of 1\% and
with a total of
$10^3$ measured neutrino events. The energy released to the recoil
electron would be about 400~keV (the reaction threshold is about
450~keV), so the average neutrino energy would
be measured to an accuracy of about 0.1~keV.
With the experimental parameters assumed, a calibrated $^{81}$Br detector
could measure the central temperature of the sun to an accuracy of about 10\%.

The requirements for a practical experiment may
be achievable since solar neutrino detectors currently under
development are designed to detect several thousand events per year,
although not yet with the energy resolution required to measure $\Delta$.
It might be possible to calibrate the
solar results by studying an intense laboratory source of
$^7$Be neutrinos with the same detector as used in the solar
observations\cite{Drukier84}.

This work was supported by NSF grant \#PHY92-45317.

\begin{figure}
\caption[]{The Energy Profile for the 862~keV line.  The probability for
the emission of a neutrino with energy $q_{\rm obs}$ in the laboratory
frame is shown as a function of $q_{\rm obs} - q_{\rm lab}$,
where $q_{\rm lab} = 861.84$~keV and $q_{\rm peak} = 862.27$~keV.
The line profile was computed by
averaging the probability distribution at a fixed temperature over the
Bahcall-Pinsonneault standard solar model with helium
diffusion\cite{Bahcall88}.}
\end{figure}

\vfill\eject
\end{document}